\documentclass[12pt]{article}
\usepackage{amssymb} 
\begin{document} 

\vskip2cm

\begin{center}
{\bf PRESENT STATUS OF NEUTRINO MIXING}

\vspace{0.3cm}
S. M. Bilenky
\footnote{ A lecture at the Advanced Study Institute "Symmetries and Spin",
 Praha-Spin-2001, Czech Republic, July 15-28, 2001}\\
\vspace{0.3cm}
{\em Physik-Department, Technische Universit\"at M\"unchen,
James-Franck-Strasse,
D-85748  Garching, Germany \\} 
\vskip0.3cm
{\em Joint Institute for Nuclear Research, Dubna, Russia}\\

\end{center}
\begin{abstract}
A short review of the status of neutrino mixing and neutrino 
oscillations is given.
The basics of neutrino mixing and oscillations is
discussed. The latest evidences of neutrino oscillations obtained in 
 the Super-Kamiokande and the SNO solar neutrino
experiments and in the Super-Kamiokande atmospheric neutrino experiment 
are considered. The results of solar and atmospheric neutrino experiments are
discussed from the point of view of the three-neutrino mixing.
\end{abstract}

\section{Neutrino mixing}

Strong evidences in favor of neutrino masses and neutrino oscillations were
obtained in atmospheric \cite{AS-K} and solar
\cite{Cl,Kam,GALLEX,GNO,SAGE,S-K,SNO} 
neutrino
experiments. 
 We will discuss here
the latest experimental data and implications for 
 neutrino mixing that can be inferred from the
existing data.
\footnote{Indications in favor of $\bar \nu_{\mu} \to \bar \nu_{e}$
oscillations 
which were obtained in the accelerator LSND experiment \cite{LSND}, require
confirmation. We will not consider LSND data here.}

Investigation of neutrino oscillations is based on two fundamental
experimental facts

\begin{enumerate}
\item 
Interaction of neutrino with matter is described by the Standard
charged current (CC) and neutral current (NC) Lagrangians
\begin{equation}
\mathcal{L}_{I}^{\mathrm{CC}}=
- \frac{g}{2\sqrt{2}} \,
j^{\mathrm{CC}}_{\alpha} \, W^{\alpha}
+\mathrm{h.c.}
\,,
\label{001}
\end{equation}
and
\begin{equation}
\mathcal{L}_{I}^{\mathrm{NC}}
=
- \frac{g}{2\cos\theta_{W}} \,
j^{\mathrm{NC}}_{\alpha} \, Z^{\alpha}
\,.
\label{002}
\end{equation}

Here $g$ is the electroweak interaction constant, $\theta_W$  is weak 
(Weinberg) angle, $W^{\alpha}$ and $Z^{\alpha}$ are the fields of
$W^{\pm}$ and $Z^{0}$ vector bosons and 
\begin{equation}
j^{\mathrm{CC}}_{\alpha} = \sum_{l} \bar \nu_{lL} \gamma_{\alpha}l_{L}
\,~;\,~~j^{\mathrm{NC}}_{\alpha} =\sum_{l} \bar \nu_{lL}
\gamma_{\alpha}\nu_{lL} 
\label{003}
\end{equation}

are the leptonic charged and neutral currents.

The CC and NC interactions (\ref{001}) and (\ref{002}) conserve flavor
 lepton numbers
 and determine {\em the notion} of
flavor neutrinos and antineutrinos.
For example,  neutrino that is
produced  together with $\mu^{+}$ in the decay
 $\pi^{+}\to \mu^{+} + \nu_{\mu}$ is the left-handed muon neutrino $\nu_{\mu}$,
antineutrino that is produced together with electron in the decay
$n \to p + e^{-} + \bar \nu_{e}$ is the right-handed electron antineutrino
$\bar \nu_{e}$
etc.\\ 
\item Three flavor neutrinos exist in nature. 
The number of the flavor neutrinos $n_{\nu_{f}}$ was determined from experiments
on the measurement of the width of the decay $Z \to \nu_{l}+ \bar\nu_{l}$
(SLC, LEP).
In the LEP experiments it was obtained that
\begin{equation}
n_{\nu_{f}} = 3.00 \pm 0.06\,.
\label{004}
\end{equation}
\end{enumerate}

According to \emph {the neutrino mixing hypothesis} \cite{P,MNS}
 masses of neutrinos are
different from zero and
fields of massive neutrinos $\nu_{i}$ enter into the CC and the NC
Lagrangians 
(\ref{001}) and (\ref{002})
in the mixed form
\begin{equation}
\nu_{{l}L}
=
\sum_{i}
U_{{l}i}
\,
\nu_{iL}
\label{005}
\,,
\end{equation}
where $\nu_i$ is the field of neutrino with mass $m_i$
and U is the unitary PMNS mixing matrix.

The relation (\ref{005}) leads to a violation of flavor lepton numbers. 
The effects of the violation of flavor lepton numbers can be revealed in 
neutrino oscillation experiments.
 We will
discuss neutrino oscillation experiments later.
 Now we will consider different
general possibilities of the neutrino mixing (see, for example \cite{BiP,BGG})

If neutrino masses are different from zero,
 there is a neutrino mass term in the
total Lagrangian. The structure of the mass term depends on the mechanism of
neutrino mass generation.
Only left-handed neutrino fields $\nu_{lL}$
enter into the Lagrangian of the weak interaction (\ref{001}) and (\ref{002}). 
In the neutrino mass term both
 $\nu_{lL}$
 and singlet $\nu_{lR}$ fields can enter. If $\nu_{lL}$ and
$\nu_{lR}$  enter into the
 mass term in such a form that
the total lepton number $L$ is conserved,
 in this case fields of massive neutrinos  are four-component
 Dirac fields and
neutrino $\nu_{i}$
and antineutrino $ \bar\nu_{i}$ have opposite lepton numbers. 
The corresponding mass term is called the Dirac mass term.
The number of the 
massive neutrinos in the case of the standard Dirac mass term is
equal
 to the number of flavor neutrinos.
The Dirac mass term can be generated
 by the Standard Higgs mechanism with a Higgs doublet.

If the lepton number is not conserved,
only left-handed components $\nu_{lL}$ can enter into the neutrino mass term.
The corresponding mass term is called the Majorana mass term.
It is a product of left-handed components $\nu_{lL}$  and right-handed
 components $(\nu_{lL})^c $, determined by the relation

\begin{equation} 
(\nu_{lL})^c =C \bar\nu_{lL}^T\,,
\label{006a}
\end{equation}
 where
 $C$ is the matrix of the charge conjugation that satisfies the conditions
 
\begin{equation} 
C\gamma_{\alpha}^{T}C^{-1}= -\gamma_{\alpha}\,;~~C^{T}=-C\,.
\label{006b}
\end{equation}

In the case of the Majorana mass term the fields  $\nu_i$ in (\ref{005})
are two-component Majorana fields that satisfy the condition
\begin{equation}
\nu_{i} = \nu_{i}^{c}\,.
\label{006}
\end{equation}

The condition (\ref{006}) means that  neutrinos and 
antineutrinos, quanta of the Majorana 
field $\nu_{i}$, are identical particles.
The number of the massive neutrinos in the case of the Majorana mass term 
is equal to three.
The Majorana mass term requires a beyond the Standard Model mechanism of 
neutrino mass
generation 
with Higgs triplets.

In the more general case  both $\nu_{lL}$ and $\nu_{lR}$ fields
enter into the mass term and there are no conserved
 lepton numbers (the Dirac and Majorana mass term).
Because the lepton numbers are not conserved there is no possibility to
distinguish neutrino and antineutrino and  
fields of neutrinos with definite masses $\nu_i$ 
in the case of the Dirac and Majorana mass term 
are two-component Majorana fields. If three 
left-handed fields $\nu_{lL}$ and three right-handed fields $\nu_{lR}$
 enter into the mass term,
the number of massive Majorana neutrinos is equal to 6.
For the mixing we have 
\begin{equation}
\nu_{lL}
=
\sum_{i=1}^6 U_{li} \nu_{iL}
\qquad
(l=e,\mu,\tau)
\label{006c}
\end{equation}
and 

\begin{equation}
(\nu_{lR})^c
=
\sum_{i=1}^6 U_{{\overline l}i} \nu_{iL}\,,
\label{007}
\end{equation}
where $U$ is the $ 6\times 6 $
unitary matrix and the fields
$\nu_i$ satisfy the condition (\ref{006}).

In the framework of the Dirac and Majorana mass term there exist a
plausible mechanism of neutrino mass generation,
which is called the see-saw mechanism
\cite{see}.
This mechanism is based on the assumption that lepton number is violated
by the right-handed Majorana mass term at the scale $ M$, which  
is much larger than the electroweak scale $\simeq 300$ GeV.
In the see-saw case  
in the spectrum of 
masses of Majorana particles there are three light masses 
$m_{k}$ (masses of neutrinos)
and three very heavy masses $M_{k}\simeq M$ (k=1,2,3).
Masses of neutrinos are connected with the masses of the heavy Majorana
particles by the see-saw relation
\begin{equation}
m_k  \simeq \frac {(m_{k}^{\mathrm{f}})^2} {M_k} <<  m_{k}^{\mathrm{f}}
\label{009}
\,.
\end{equation}
where $m_{k}^{\mathrm{f}}$ is the mass of lepton or quark in $k$-family.
The see-saw mechanism 
connects the smallness of neutrino masses with respect to the masses of all
other fundamental fermions
with a new physics at a large scale.

The fields $\nu_{lR}$ do not enter into
 the Lagrangian of the standard electroweak
 interaction and are called sterile. The
 nature and the number of sterile
 fields depend on model. They can be not
 only singlet right-handed neutrino 
 fields but also SUSY fields and so on. 
Thus, in the most general case for the mixing we have
\begin{equation}
\nu_{lL}
=
\sum_{i=1}^{3+n_s} U_{li} \nu_{iL}
\label{010}
\end{equation}
and 
\begin{equation}
\nu_{sL}
=\sum_{i=1}^{3+n_s} U_{si} \nu_{iL}\,,
\label{011}
\end{equation}

where $n_{s}$ is the number of sterile fields and $U$ is a
 $(3+n_{s}) \times(3+n_{s}) $ unitary matrix. 

\section{Neutrino oscillations}

Neutrino oscillations  is the most important consequence 
of neutrino mixing.  
Neutrino oscillations were first considered by B. 
Pontecorvo in 1957-58 \cite{P}. It was very courageous conjecture, made
 at the time when
 two-component theory for {\em massless} neutrino was
proposed \cite{Dau} and it was established in the Goldhaber et al
 experiment \cite{Gold} that neutrino is left-handed particle.           
 Only 
electron neutrino was known at that time. B.Pontecorvo considered
transition of $\nu_{e}$ into sterile left-handed state $\bar\nu_{eL}$

If there is neutrino mixing 
and neutrino mass- squared differences 
are much smaller than  square of neutrino energy, 
the normalized state of flavor (sterile) neutrino with
 momentum $\vec{p}$  
is given by  
\begin{equation}
|\nu_\alpha\rangle
=
\sum_{i} U_{{\alpha}i}^* \,~ |\nu_i\rangle
\,.
\label{012}
\end{equation}

Here $|\nu_i\rangle$ is the vector of the state of neutrino with mass $ m_{i}$,
momentum $\vec{p}$, energy

\begin{equation}
E_i
=
\sqrt{p^2 + m_i^2 }
\simeq
p + \frac{ m_i^2 }{ 2 p }
\label{013}
\end{equation}
and  negative helicity 
(up to the terms  $\frac {m_{i}^2} {p^2}$) 

Thus, if there is neutrino mixing, the state of flavor (sterile) neutrino
is a {\em superposition} of states of neutrinos with definite masses.
The phenomenon
of neutrino oscillations is based on the  relation (\ref{012}).
This relation is similar to the well known relations 
that connect the states of
$K^0$ and $\bar K^0$ mesons with the states of $K_{S}$ and  $K_{L}$ mesons,
 particles with
definite masses and times of life.

Let us consider now the evolution in vacuum of a mixed state given by 
Eq. (\ref{012}).
 If
at the initial time t=0 the state of neutrino is  $|\nu_\alpha\rangle$, 
at the time $t$ for the neutrino state we have 

\begin{equation}
|\nu_\alpha\rangle_t
=
\sum U_{{\alpha}i}^* \,~ e^{ - i E_i t } \,~ |\nu_i\rangle
=
\sum_{\alpha'}A_{\nu_{\alpha} \to \nu_{\alpha'}}(t)\,~ |\nu_{\alpha'}\rangle\,,
\label{014}
\end{equation}

where
\begin{equation}
 A_{\nu_{\alpha} \to \nu_{\alpha'}}(t)
= \sum_i U_{\alpha' i}\,~e^{-
iE_it}\,~U_{\alpha i}^*
\label{015}
\end{equation}
is the amplitude of the transition $\nu_\alpha \rightarrow 
\nu_{\alpha'}$
during the time $t$.

The expression (\ref{015}) for the transition amplitude
has a 
simple meaning. The term 
$U^*_{\alpha i}$ is 
the amplitude of the transition from the state  $|\alpha\rangle$
to the state  $|i\rangle$; the term  $e^{-iE_it}$
describes evolution in the state with energy $E_i$; the term 
$U_{\alpha' i}$
is the amplitude of the transition from the state   
$|i\rangle$ to the state   $|\alpha'\rangle$.

It follows from  (\ref{015}) that transition between different neutrinos
is an effect of neutrino masses differences and neutrino mixing.
In fact, if all neutrino masses  in
 (\ref{015}) are the same and/or $U=1$ in this case
$ A_{\nu_{\alpha} \to \nu_{\alpha'}}(t) = e^{-iE t}\,\delta_{\alpha'\,\alpha}$
and there are no transitions between different types of neutrinos in 
a neutrino beam.

Let us enumerate neutrino masses in such a way that $m_1<m_2<...$.
 From (\ref{015}) for the 
probability of the transition $\nu_\alpha\to\nu_{\alpha'}$
in vacuum we have the following expression
\begin{equation}
P(\nu_\alpha\to\nu_{\alpha'})
=
\left|
\sum_{i} U_{{\alpha'} i} \,~ U_{{\alpha}i}^*
 e^{ - i
 \,~ 
\Delta{m}^2_{i1} \frac {L}{2 E} }
\right|^2
\,,
\label{016}
\end{equation}
where $\Delta{m}^2_{i1}=m^{2}_{i} - m^{2}_{1}$ and $L \simeq t$ is the distance
between a neutrino source and a neutrino detector and $E$ is neutrino energy.

The simplest case of neutrino oscillations is the oscillations between two types
of neutrinos. In this case  
the index $i$ in (\ref{016}) takes values 1 and 2. Taking into account
the unitarity relation
\begin{equation}
 U_{{\alpha'} 1} \, U_{{\alpha}1}^* =\delta_{\alpha\alpha'}-
U_{{\alpha'} 2} \, U_{{\alpha}2}^*
\label{017}
\end{equation}
for the 
transition probability we have

\begin{equation}
{\mathrm P}(\nu_\alpha \to \nu_{\alpha'}) =
|\delta_{{\alpha'}\alpha} + U_{\alpha'_2}  U_{\alpha_2}^*
\,~ (e^{- i \Delta m^2_{2 1} \frac {L} {2E}} -1)|^2 \,,
\label{018}
\end{equation}
where $\Delta m^2_{2 1}\equiv \Delta m^2 $. The neutrino mixing matrix for the
$2\times 2 $ case has the following general form

\begin{eqnarray}
\left(
\begin{array}{cc} \displaystyle
\cos \theta
\null & \null \displaystyle
\sin \theta
\\ \displaystyle
-\sin \theta
\null & \null \displaystyle
\cos \theta 
\end{array}
\right)\,,
\label{019}
\end{eqnarray}

where $\theta $ is the mixing angle.

For  $\alpha' \neq \alpha$ from (\ref{018})we have

\begin{equation}
{\mathrm P}(\nu_\alpha \to \nu_{\alpha}')  
= \frac {1} {2} {\mathrm A}_{{\alpha'};\alpha}\,~ (1 - \cos \Delta m^2 \frac 
{L} {2E})\,.
\label{020}
\end{equation}

Here

\begin{equation} 
{\mathrm A}_{\alpha'; \alpha} = 4 |U_{{\alpha'}2}|^2  
|U_{{\alpha}2}|^2 = \sin^{2} 2\theta\,.
\label{021}
\end{equation}

For the survival probability from Eq. (\ref{018}) 
we have

\begin{equation}
{\mathrm P}(\nu_\alpha \to \nu_\alpha) =
 1 - \frac {1} {2}{\mathrm B}_{\alpha ; \alpha}\,~ (1 - \cos \frac {\Delta m^2 L} {2E})\,,
\label{022}
\end{equation}

where 
\begin{equation}
{\mathrm B}_{\alpha ; \alpha}= 4 |U_{{\alpha}2}|^2\,~ ( 1 - |U_{{\alpha}2}|^2)=
\sin^{2} 2\theta
\label{023}
\end{equation}

Thus, in the case of transitions between two types of neutrinos
all transition
probabilities are characterized by the two oscillation 
parameters  $sin^2 2 \theta$ and 
$\Delta m^2$. Notice that due to the unitarity of the mixing matrix
in the two-neutrino case ${\mathrm A}_{{\alpha'};\alpha}= 
{\mathrm B}_{\alpha ; \alpha}$.
The formulas (\ref{020}) -(\ref{023}) are the standard ones. They are 
usually used for an analysis of experimental data.

From (\ref{020}) and (\ref{022})
it is obvious that neutrino oscillations can be observed if the condition
\begin{equation}
 \frac {\Delta m^2 L} {2E} \gtrsim 1
\label{024}
\end{equation}
is satisfied. We can rewrite this condition in the form

\begin{equation}
 \cdot\frac {\Delta m^2(\rm{eV}^{2}) L(\rm{m})} {E(\rm{MeV})} \gtrsim 1
\label{025}
\end{equation}

The values of the parameter $\frac {L} {E}$
depend on conditions of an experiment. The larger 
$\frac {L} {E}$ the more sensitive a neutrino oscillation 
experiment to the presumably
small values
of  $\Delta m^{2}$. The average values of the parameter $\frac {L} {E}$ 
 for experiments with accelerator neutrinos, 
 reactor antineutrinos, atmospheric neutrinos and solar neutrinos are,
 correspondingly, in the ranges $10^{-1} - 10^3$ , $10^2 - 10^3$, $10 - 10^4$  
and $10^{10} - 10^{11}$. 
Thus, neutrino oscillation experiments are  
sensitive to the parameter
 $\Delta m^{2}$ in a wide range from $\Delta m^{2}\simeq 10 \,~\rm{eV}^{2}$
to  $\Delta m^{2}\simeq 10^{-11}\,~ \rm{eV}^{2}$.

\section{Neutrino oscillation experiments}
There exist at present compelling evidence in favor of neutrino oscillations
that were obtained in all solar and atmospheric neutrino experiments.
We will discuss here only the latest results of the  Super-Kamiokande 
\cite{S-K} and SNO \cite{SNO}
 solar neutrino experiments and the result of the Super-Kamiokande
atmospheric neutrino experiment \cite{AS-K}.

\subsection{Solar neutrinos}

The energy of the sun is generated in the reactions of the thermonuclear 
pp and CNO cycles. From 
thermodynamical
point of view
the energy of the sun is produced in the transition
\begin{equation}
4 \, p + 2 \, e^-
\to
{}^4He + 2 \, \nu_e 
\,,
\label{026}
\end{equation}
in which the energy

\begin{equation}
Q = 4 m_p + 2 m_e - m_{^4He} \simeq 26.7 MeV 
\label{027}
\end{equation}
is released.
From (\ref{026}) we can easily obtain a model independent relation
\begin{equation}
\int {\frac{1}{2} (Q -2 E)\,~\Phi_{\nu_{e}}^{tot}(E)  dE} =
\frac{L_\odot} {4 \pi R^2}\,,
\label{028}
\end{equation}   
which connect the luminosity of the sun $L_\odot$
 with the initial total
flux of the solar electron neutrinos 

\begin{equation}
\Phi_{\nu_{e}}^{tot}(E) =\sum_i \Phi_{\nu_{e}}^{i}(E)\,,
\label{029}
\end{equation}
where $ \Phi_{\nu_{e}}^{i}$ is the flux of $\nu_{e}$ from the source i.
In Eq. (\ref{028}) $R$ is the distance between the sun and the earth .

The main source of solar neutrinos is $pp$ reaction
\begin{equation}
p+p \to d + e^- + \nu_{e}\,.
\label{030}
\end{equation}
Low energy neutrinos with
the maximum neutrino energy  0.42 MeV 
is produced in this reaction.
The total
flux of the $pp$ neutrinos, predicted by the Standard Solar Model BP2000(SSM) 
\cite{BP}
and determined mainly by the luminosity relation (\ref{028}), is equal to 
$\Phi_{\nu_{e}}^{pp}= 5.94 \cdot 10^{10}cm^{-2}s^{-1}$

The next important source is the reaction
\begin{equation}
e^{-} + ^7 Be \to ^7 Li +\nu_{e}\,. 
\label{031}
\end{equation}

In this reaction monochromatic neutrinos with energy 0.86 MeV are produced. 
The flux of
$^7 Be$ neutrinos, predicted by the SSM,  is given by  
$\Phi_{\nu_{e}}^{^7 Be}= 4.8 \cdot 10^{9}cm^{-2}s^{-1}$

In the Super-Kamiokande (S-K) and the SNO experiments 
because of high energy thresholds
mainly high energy neutrinos from the decay  
\begin{equation}
^8 B \to ^8 Be^{*} +e^{+}+\nu_{e}\,. 
\label{032}
\end{equation}
are detected. 
The maximum energy of the $^8 B $ neutrinos is equal to 15 MeV
and the flux, predicted by the SSM, is given by
$\Phi_{\nu_{e}}^{^8 B}= 5.1 \cdot 10^{6}cm^{-2}s^{-1}$.

In the S-K experiment large 
water Cherenkov
detector is used (50 ktons of $\rm{H}_{2}\rm{O}$). The solar neutrinos 
are detected by the observation of the elastic (ES) neutrino-electron 
 scattering
\begin{equation}
\nu_x + e \to \nu_x + e \,.
\label{033}
\end{equation}

All flavor neutrinos 
$\nu_e$, $\nu_\mu$ and $\nu_\tau$ are detected in the S-K experiment. However,
the sensitivity to $\nu_{\mu}$ and $\nu_\tau$
is much lower than sensitivity to $\nu_e$: 
the  cross 
section
of NC $\nu_{\mu}$ ($\nu_{\tau}$)  - $e$ scattering is about six times smaller 
than the cross section of CC+NC $\nu_e-e$ scattering.

During 1258 days of running, in the S-K experiment 
$18464 \begin{array}{c} +677 \\-590\end{array}$
events with energy of the recoil electrons larger than 5 MeV was observed.

The total ES event rate 
is given by

\begin{equation}
R^{ES}=  <\sigma_{\nu_{e}e}>\Phi_{\nu_{x}}^{ES}\,,
\label{034}
\end{equation}
where
\begin{equation}
\Phi_{\nu_{x}}^{ES}= \Phi_{\nu_{e}}^{ES} + \frac{<\sigma_{\nu_{\mu}e}>}
{<\sigma_{\nu_{e}e}>}\Phi_{\nu_{\mu,\tau}}^{ES}\,.
\label{035}
\end{equation}
Here  $\Phi_{\nu_{e}}^{ES}$ ($\Phi_{\nu_{\mu,\tau}}^{ES}$) is the 
the flux of solar $\nu_{e}$ ($ \nu_{\mu,\tau}$) on the earth and
$<\sigma_{\nu_{e}e}>$ and $<\sigma_{\nu_{\mu}e}>$ are the cross sections of the
processes $\nu_{e}e \to\nu_{e}e$ and $\nu_{\mu}e \to\nu_{\mu}e$,
averaged over initial spectrum of $^8 B$ neutrinos. 
Notice that the spectrum of neutrinos from the decay (\ref{032}) 
is determined by the weak interaction and is known.
We have
\begin{equation}
\frac{<\sigma_{\nu_{\mu}e}>}{<\sigma_{\nu_{e}e}>} \simeq 0.154\,.
\label{036}
\end{equation}

From the data of the S-K experiment it was obtained

\begin{equation}
(\Phi_{\nu_{x}}^{ES})_{SK}= (2.32 \pm 0.03 \pm 0.08) 
\cdot 10^{6}cm^{-2}s^{-1} \,,
\label{037}
\end{equation}

The S-K data alone do not allow to obtain an information 
separately on the fluxes of $\nu_{e}$
and
$\nu_{\mu,\tau}$ on the earth.
It became possible only after 
the data
of the SNO experiment \cite{SNO} appeared .

In the SNO experiment heavy water Cherenkov detector is used (1 kton of
$\rm{D}_{2}\rm{O}$). Solar neutrinos were detected in the experiment via the
observation
of CC reaction
\begin{equation}
\nu_e + d \to e^{-}+ p +p
\label{038}
\end{equation}
and elastic scattering (ES) reaction
\begin{equation}
\nu_{x} + e \to \nu_{x} + e\,.
\label{039}
\end{equation}

The electron kinetic energy threshold in the experiment was  6.75 MeV.
From November 1999 till January 2001 it was observed $975.4 \pm 39.7$  
CC events and $106.1 \pm 15.2$ ES events. 
For the effective flux of ES events it was obtained the value

\begin{equation}
(\Phi_{\nu_{x}}^{ES})_{SNO}=( 2.39 \pm 0.34 \pm 0.16) \cdot 10^{6}\,~
cm^{-2}s^{-1} \,,
\label{040}
\end{equation}

which is in agreement with the S-K value (\ref{037}).

The CC events rate is given by
\begin{equation}
R^{CC}=  <\sigma_{\nu_{e}d}>\Phi_{\nu_{e}}^{CC}\,,
\label{041}
\end{equation}
where $<\sigma_{\nu_{e}d}>$
is the cross section of the process (\ref{038}) averaged over initial spectrum of
the $^8 B$ neutrinos and $\Phi_{\nu_{e}}^{CC}$ is the flux of $\nu_{e}$ on
 the earth. In the SNO experiment it was found

\begin{equation}
\Phi_{\nu_{e}}^{CC} = (1.75 \pm 0.07 \pm 0.12 \pm 0.05 (\rm{theor}))
 \cdot 10^{6}\,~
cm^{-2}s^{-1} \,.
\label{042}
\end{equation}

Let us compare now (\ref{037}) and (\ref{042}). For the fluxes of  $\nu_{e}$ on
the earth in Eq. (\ref{035}) and Eq.(\ref{041}) we have, respectively

\begin{equation}
\Phi_{\nu_{e}}^{CC} = < P (\nu_{e}\to\nu_{e})>_{CC} \Phi_{\nu_{e}}^{0}\,;~~
\Phi_{\nu_{e}}^{ES} = < P (\nu_{e}\to\nu_{e})>_{ES} \Phi_{\nu_{e}}^{0}\,,
\label{043}
\end{equation}
where $\Phi_{\nu_{e}}^{0}$ is the total initial flux of the $^8 B$ neutrinos.

If the $\nu_{e}$ survival probability depends on energy, the average
 probabilities
$ < P (\nu_{e}\to\nu_{e})>_{CC}$ and $ < P (\nu_{e}\to\nu_{e})>_{ES}$ are
in principle different. \footnote{ It was shown in \cite{Villante}
 that it is possible to choose
the S-K and the SNO thresholds in such a way that these quantities
will be practically equal at any $ P (\nu_{e}\to\nu_{e})$. }.
 However, no indications of
significant energy dependence of the $\nu_{e}$ survival probability in 
the S-K and SNO energy ranges were obtained. 
In fact, in both experiments spectra of electrons
were measured. If survival probability is a constant the shapes of the spectra 
can be predicted in a model independent way. 
No sizable deviations from the predicted
spectra were found in both experiments.
Thus, we have
\begin{equation}
\Phi_{\nu_{e}}^{CC} \simeq
\Phi_{\nu_{e}}^{ES}\,.
\label{044}
\end{equation}

Taking into account this relation from 
(\ref{035}), (\ref{037}) and (\ref{042}) for the flux of $\nu_{\mu}$ and 
$\nu_{\tau}$ on the earth we obtain
\begin{equation}
\Phi_{\nu_{\mu,\tau}}^{ES}= (3.69 \pm 1.13)  \cdot 10^{6}cm^{-2}s^{-1} 
\label{045}
\end{equation}
Thus, the results of the SNO and the S-K experiments give us
the first model independent evidence ( at $\simeq 3\sigma $ level)
of the presence of $\nu_{\mu}$ and $\nu_{\tau}$ in the flux of solar neutrinos
on the earth.
The flux of $\nu_{\mu}$ and $\nu_{\tau}$  is approximately two times
 larger than the flux $\nu_{e}$.

From (\ref{042}) and (\ref{045}) for the total flux of all flavor neutrinos 
on the earth we have

\begin{equation}
\Phi_{\nu_{e,\mu,\tau}}= (5.44 \pm 0.99 ) \cdot 10^{6}cm^{-2}s^{-1} 
\label{046}
\end{equation}

This value is in a  agreement with the total flux of $^8 B$ neutrinos

\begin{equation}
\Phi_{\nu_{e}}^{SSM}= 5.05 \cdot 10^{6}cm^{-2}s^{-1}\,, 
\label{047}
\end{equation}
predicted by the SSM BP 2000 \cite{BP}.

The data of all solar neutrino experiments can be described if we assume that
two-neutrino oscillations, which are
 characterized by the two parameters $\Delta m^{2}_{sol}$ and
$\tan^{2}\theta_{sol}$, take place. 
From the analysis of all existing data, 
made under the assumption that initial fluxes are given by the SSM, 
the several
allowed regions (solutions) in the plane of these parameters  were obtained.
After the new S-K and SNO data were obtained
the large mixing angle MSW allowed regions (LMA, LOW)
became the preferable ones (see \cite{Bahcall,Fogli})
 For the best-fit values of the
oscillation parameters in the LMA region in ref. \cite{Bahcall}
it was found 
\begin{equation}
 \Delta m^{2}_{sol} = 4.5 \cdot 10^{-5}\,\rm{eV}^{2};\,~~
 \tan^{2}\theta_{sol} = 4.1 \cdot 10^{-1}\,.
\label{048}
\end{equation}

\subsection{Atmospheric neutrinos}

The decays of charged pions

\begin{equation}
\pi^{+} \to \mu^{+}+ \nu_\mu \,~~\pi^{-} \to \mu^{-}+ \bar\nu_\mu\,, 
\label{049}
\end{equation}
produced in interaction of cosmic rays with the atmosphere,
and subsequent decays of muons 

\begin{equation}
\mu^{+} \to e^{+}+ \nu_{e} +  \bar\nu_\mu\,~~
\mu^{-} \to e^{-}+ \bar\nu_{e} + \nu_\mu\
\label{050}
\end{equation}
are the main source of the atmospheric neutrinos.

In the Super-Kamiokande experiment muons and electrons,
 produced in interaction of the atmospheric
$\nu_{\mu}$ and $\nu_{e}$ with nuclei, are detected in the
the large water Cherenkov detector. 
The first compelling evidence in favor of neutrino oscillations 
was obtained by the S-K collaboration in 1998.

For the high-energy neutrinos the distance $L$ between 
the region where neutrinos
are
produced and the detector is determined by the zenith angle $\theta_{z}$. 
Down-going neutrinos 
($\cos \theta_{z} = 1$) pass the distance about 20 km. Up-going neutrinos 
($\cos\theta_{z} = -1$)  pass the distance about 13000 km.
If there is no neutrino oscillations for the number of muon (electron) events
we have

\begin{equation}
N_{l}(\cos\theta_{z})= N_{l}( -\cos \theta_{z})\,~~ (l=e,\mu) \,.
\label{051}
\end{equation}
 
The S-K collaboration measured $\cos \theta_{z}$ dependence of the number
of electron and muon events. For the electron events  
no $\cos \theta_{z}$ asymmetry was observed. The data are in a good agreement
with the Monte Carlo prediction, obtained under the assumption 
of no oscillations.
For the muon neutrinos in the Multi-GeV region (neutrinos with 
energies larger than 1.3 GeV) 
strong $\cos \theta_{z}$ asymmetry was observed.
For the ratio of the total numbers of up-going and down-going 
high-energy muons 
it was found \cite{AS-K}

\begin{equation}
\left(\frac{U}{D}\right)_{\mu} = 0.54 \pm 0.04 \pm 0.01
\label{052}
\end{equation}
The data of the S-K atmospheric neutrino experiment 
can be described if we assume that $\nu_\mu \to \nu_\tau$ oscillations
 take place. From the data of the S-K experiment the
following best-fit values of the oscillation parameters were found \cite{SKatm}

\begin{equation}
\Delta m^{2}_{atm} \simeq 2.5 \cdot 10^{-3} \rm{eV}^2 \,;~~~
\sin^2 2\theta_{atm} \simeq 1
\label{053}
\end{equation}

\section{ Oscillations of solar and atmospheric neutrinos 
from the point of view of
three-neutrino mixing}

The data of all solar and atmospheric neutrino oscillation experiments are
described by the two-neutrino oscillations.
We 
 will discuss here the origin of such a picture in the 
framework of the minimal scheme of the mixing of
three massive neutrinos (see, for example \cite{BGG}).

The probability of the transition $\nu_{\alpha} \to \nu_{\alpha'}$  
in vacuum is given by the general expression (\ref{016}). 
Taking into account (\ref{048}) and (\ref{053}),
 we will assume that
the following hierarchy relation holds

\begin{equation}
\Delta m^{2}_{21} \ll \Delta m^{2}_{31}\,.
\label{054}
\end{equation}

Let us consider first neutrino oscillations in atmospheric 
and long baseline (LBL) reactor and  accelerator
experiments. 
In these experiments

\begin{equation}
\Delta{m}^2_{21} \frac {L}{2 E}\ll 1
\label{055}
\end{equation}
and we can neglect the contribution of 
$\Delta m^{2}_{21}$ to the transition probability (\ref{016}).

Taking into account 
the unitarity relation 
\begin{equation}
\sum_{i=1,2} U_{{\alpha'} i} \, U_{{\alpha}i}^* =\delta_{\alpha\alpha'}-
U_{{\alpha'} 3} \, U_{{\alpha}3}^*
\label{056}
\end{equation}
for the transition probability in the leading approximation 
we obtain the following relation
\begin{equation}
P (\nu_\alpha\to\nu_{\alpha'})
\simeq
\left|
\delta_{\alpha \alpha'}
+
 U_{{\alpha'} 3} \, U_{{\alpha}3}^*
\left( e^{ - i
 \, 
\Delta m^{2}_{31} \frac {L}{2 E} } - 1 \right)
\right|^2\,.
\label{057}
\end{equation}

Thus, if inequality (\ref{055}) is satisfied, the  probabilities 
of transition $\nu_\alpha\to\nu_{\alpha'}$
in
the atmospheric (LBL) experiments are determined by
$\Delta m^{2}_{31}\equiv \Delta m^{2}_{atm}$
 and the elements of the mixing matrix $U_{{\alpha}3}$,
which connect flavor neutrinos with the heaviest neutrino
$\nu_{3}$.

For  $\alpha' \neq \alpha$ from (\ref{057}) we have

\begin{equation}
 P(\nu_{\alpha} \to \nu_{\alpha'}) =
\frac {1} {2} {\mathrm A}_{{\alpha'};\alpha}\,~ (1 - \cos\, \Delta m^{2}_{31} \frac 
{L} {2E})\,,
\label{058}
\end{equation}

where
\begin{equation} 
{\mathrm A}_{\alpha'; \alpha} = 4 |U_{{\alpha'}3}|^2  
|U_{{\alpha}3}|^2\,.
\label{059}
\end{equation}

For the survival probability from 
Eq. (\ref{057}) we obtain 

\begin{equation}
P(\nu_\alpha \to \nu_\alpha) 
\simeq 1 - \frac {1} {2} {\mathrm B}_{\alpha; \alpha}\,~
(1 - \cos\,~ \Delta m^2_{3 1} \frac {L} {2E})\,,
\label{060}
\end{equation}

where 
\begin{equation}
{\mathrm B}_{\alpha; \alpha} =
 4\,~ |U_{{\alpha}3}|^2 (1 -  
|U_{{\alpha}3}|^2)\,.
\label{061}
\end{equation}
Thus, due to the hierarchy (\ref{054}) oscillations of atmospheric  
(LBL) neutrinos are described by the two-neutrino type formulas with
the same $\Delta m^2_{3 1}$ for all channels. The quantities 
${\mathrm A}_{\alpha'; \alpha}$ and ${\mathrm B}_{\alpha ;\alpha}$
are oscillation amplitudes. From the unitarity of the mixing matrix it follows
that they
are connected by the relation

\begin{equation} 
\sum_{{\alpha} \not={\alpha}'} {\mathrm A}_{\alpha'; \alpha}=
 {\mathrm B}_{\alpha; \alpha}    
\label{062}
\end{equation}

and satisfy the inequalities

\begin{equation}
 0 \leq {\mathrm B}_{\alpha; \alpha} \leq 1;\,~
0\leq 
 {\mathrm A}_{\alpha'; \alpha}\leq 1
\label{063}
\end{equation}
The oscillation amplitudes depend on the two parameters
$|U_{{\mu}3}|^2$ and $|U_{{\tau}3}|^2 $ (due to the 
unitarity of the mixing matrix
 $|U_{e3}|^2=1 -
|U_{{\mu}3}|^2 -|U_{{\mu}3}|^2$ ).

It is important to stress that
the phase of the matrix elements $U_{{\alpha}3}$ does not enter
 into expression (\ref{058}) for the transition probability.
Thus, if there is hierarchy (\ref{054}), the relation
\begin{equation}
 P(\nu_{\alpha} \to \nu_{\alpha'}) =
 P(\bar \nu_{\alpha} \to \bar \nu_{\alpha'}) 
\label{064}
\end{equation}
is satisfied automatically 
and (in the leading approximation) CP violation 
in the lepton sector can not be revealed 
by the investigation
of neutrino oscillations in LBL (atmospheric) neutrino experiments.

 {\em The hierarchy of neutrino mass squared differences (\ref{054})
 is the reason
why in the leading approximation the results of the atmospheric
 and the LBL neutrino oscillation experiments can be described by the standard
two- neutrino formulas}.

Let us consider now solar neutrinos. The probability of solar $\nu_{e}$
to survive in vacuum is given by the expression
\begin{equation}
P^{sol}(\nu_e\to\nu_e)
=
\left|
\sum_{i=1,2}| U_{e i}|^{2} \, 
e^{ - i \, 
\Delta{m}^2_{i1} \frac {L}{2 E} } 
+| U_{e 3}|^{2}e^{ - i \, 
\Delta{m}^2_{31} \frac {L}{2 E} }\right|^{2}\,.
\label{065}
\end{equation}
We are interested in 
the survival probability averaged          
over the region, where neutrinos are
produced, over neutrino energies etc.
Because of the hierarchy (\ref{054}) 
 the interference between the
 first and the second term in (\ref{065}) disappears
due to averaging  and
 for the averaged survival probability we have
\begin{equation}
P^{sol}(\nu_e\to\nu_e)
=| U_{e 3}|^{4}
+(1 -| U_{e 3}|^{2})^{2}\,~ P^{1,2}(\nu_e\to\nu_e)\,,
\label{066}
\end{equation}

where $ P^{1,2}(\nu_e\to\nu_e)$ is two-neutrino survival probability
which depend on $\Delta{m}^2_{21}$
and the angle $\theta_{12}$ that is determined by the relations
\begin{equation}
\cos^2\theta_{21}
=
\frac{ |U_{e1}|^2 }{\sum_{i=1,2}|U_{ei}|^2 }
\,,
~~~
\sin^2\theta_{21}
= 
\frac{ |U_{e2}|^2 }{\sum_{i=1,2}|U_{ei}|^2 }
\,.
\label{067}
\end{equation}
 
It was shown \cite{Schramm} that the relation 
(\ref{066}) is valid also in the case of matter.
In this case the electron density $\rho_{e}$ in the  
effective matter potential must be replaced by
 $(1 -| U_{e 3}|^{2})\rho_{e}$.

From data  of the long baseline reactor
experiments CHOOZ \cite{CHOOZ} and Palo Verde \cite{PaloV} and from the 
data of the S-K 
atmospheric neutrino experiment \cite{AS-K}
it follows that the element $| U_{e 3}|^{2}$
is small.
The best limit on  $| U_{e 3}|^{2}$ can be obtained from
the results of the CHOOZ experiment.
 In this
experiment  $\bar \nu_{e}$'s from two reactors at the distance of about 1 km
from the detector were detected. No indications in favor 
neutrino oscillations
were found. 

The data of the CHOOZ experiment were analyzed 
in ref. \cite{CHOOZ}
under the assumption of two-neutrino oscillations and the exclusion
plot in the plane of the parameters $\sin^{2}2 \theta \equiv {\mathrm B}_{e;e}$
and $\Delta m^{2}\equiv \Delta m^{2}_{31} $ was obtained.
From this plot for a fixed value of $\Delta m^{2}_{31}$
we have
\begin{equation}
{\mathrm B}_{e;e}\leq {\mathrm B}_{e;e}^{0}(\Delta m^{2}_{31})\,,
\label{068}
\end{equation}

From (\ref{059}) and (\ref{068}) it follows that 

\begin{equation}
|U_{e 3}|^{2} \leq  
\frac{1}{2}\,\left(1 - \sqrt{1- B_{e;e}^{0}}\right)
\label{069}
\end{equation}
or
\begin{equation}
|U_{e 3}|^{2} \geq
\frac{1}{2}\,\left(1 + \sqrt{1-B_{e;e}^{0} }\right)\,.
\label{070}
\end{equation}

From the CHOOZ exclusion plot
we can conclude that in the region 
$\Delta m^{2}_{31} \geq 2\cdot10^{-3} \rm{eV}^{2}$,
$B_{e;e}^{0}\leq 2\cdot 10^{-2}$. If the value of $\Delta m^{2}_{31}$
lies in this region from (\ref{069}) and (\ref{070})
it follows that the element  $ |U_{e 3}|^{2}$
can be small (inequality (\ref{069})) or large
(inequality
(\ref{070})).

This last possibility is excluded by the
solar neutrino data. In fact,
if $ |U_{e 3}|^{2}$
is close to one,  
than from Eq. (\ref{066}) it is obvious that the
suppression
of the flux of solar $\nu_{e}$, observed in all solar neutrino
experiments,
cannot be explained by  neutrino oscillations. 
 Thus, from the results 
of the CHOOZ 
and solar neutrino experiments it follows that the upper bound of
 $ |U_{e 3}|^{2}$ is given by inequality (\ref{069}). 

At  
$\Delta m^{2}_{31} = 2.5\cdot10^{-3} \rm{eV}^{2}$ (the
S-K best-fit value) we have
\begin{equation}
|U_{e 3}|^{2} \lesssim
4\cdot 10^{-2}\,.
\label{071}
\end{equation}

{\em The smallness of $|U_{e 3}|^{2}$  is the reason 
why in the leading approximation oscillations of solar neutrinos are described
by the standard two-neutrino formula}. 

In the limiting case 
$|U_{e 3}|^{2}= 0$ oscillations of solar and atmospheric (LBL) neutrinos
are decoupled \cite{BG}. In this approximation  
solar neutrino experiments allow to
 obtain information on the values of the parameters $\Delta m^{2}_{21}$ and
$\theta_{12}$,
that characterize oscillations $\nu_{e} \to \nu_{\mu,\tau}$ and
the atmospheric (LBL) experiments  allow to obtain information
 on  the values of the parameters 
$\Delta m^{2}_{31}$ and
$\theta_{23}$, which characterize oscillations
 $\nu_{\mu} \to \nu_{\tau}$.
 
There is, however, no general theoretical reasons for $|U_{e 3}|^{2}$
to be equal to zero. The exact value of the parameter $|U_{e 3}|^{2}$
is of a great interest for further investigation of neutrino mixing.
If  $|U_{e 3}|^{2}$ has a nonzero value and  the 
parameter $\Delta m^{2}_{21}$
is not very small there is a possibility to investigate effects of
three-neutrino mixing and, in particular, fundamental effects of CP-violation
in the lepton sector in the future LBL experiments with neutrinos from the 
Neutrino
factories  and the Superbeam facilities 
(see \cite{NuFact,NuBeam}
and references therein).

\section{Conclusions}

About 40 years passed from the first idea of neutrino oscillations,
put forward by B. Pontecorvo in 1957-58, to the 
evidences for neutrino oscillations, obtained in the
atmospheric and solar experiments. The first idea of neutrino masses and mixing
was based on an 
analogy with $K^{0}-\bar K^{0}$ mixing and on the fact that there
is no
general
 principle (like gauge invariance in the case of photon) 
that oblige neutrino to be massless particle.
In seventies neutrino mixing was considered as a natural consequence of the
analogy between quarks and leptons. 
After the appearance of GUT and other models beyond the Standard Model 
and after the see-saw mechanism of neutrino mass generation was proposed 
neutrino masses and mixing are considered as a signature of a new physics at  
a scale much larger than the electroweak scale.

Today we have not only evidence in favor of neutrino oscillations but also
an information about the values of 
parameters, which characterize
neutrino oscillations. With many new ongoing and 
future experiments ((K2K) \cite{K2K},
KamLAND \cite{KamLAND}, BOREXINO \cite{BOREX} 
MINOS \cite{MINOS}, CNGS \cite{CNGS} and other)
evidence in favor of neutrino oscillations most probably will be more vigorous
and neutrino oscillations parameters will be determined
with much better accuracy
than today.

There exist, however, several unsolved {\em basic} problems of neutrino
 masses and mixing.
From our point of view they are
\begin{enumerate}
\item How many massive light neutrinos exist in nature?

\item Are massive neutrinos Dirac or Majorana particles?

\item What is the value of the minimal neutrino mass $m_{1}$?
\end{enumerate}

An answer to the first question probably will be obtained in the MiniBooNE 
experiment \cite{MiniB}, which will check the LSND claim.

An answer to the second question can be obtained from future experiments
on the search for neutrinoless double- $\beta $ decay (see {\cite{Klap}).

Finally, we can hope to get some answer on the third question from
the future experiment on the
investigation of the high-energy part of the $\beta $ spectrum of 
$^3 H $ \cite{KATRIN}.

Existing solar and atmospheric neutrino data are well described
by practically decoupled $\nu_{e} \to\nu_{\mu,\tau}$ and
$\nu_{\mu} \to\nu_{\tau}$ oscillations.
Detailed investigation of effects of three (or may be more?) neutrino 
masses and mixing and in particular effects of CP-violation in the lepton sector
will require such high-intensity neutrino facilities as the Superbeam
facilities and the Neutrino factories  (see \cite{NuFact,NuBeam}).

I acknowledge the Alexander von Humboldt Foundation for the support.

\end{document}